\documentclass[epsfig,twocolumn,preprintnumbers,amsmath,amssymb]{revtex4}
\usepackage{epsfig}
\usepackage{amsmath}
\usepackage{fancyhdr}

\def\eps{\epsilon}

\newcommand {\snn}       {\ensuremath{\sqrt{s_{\rm NN}}}}

\newcommand {\av}[1]     {\ensuremath{\left< #1 \right>}}
\newcommand {\abs}[1]    {\ensuremath{\left| #1 \right|}}
\newcommand {\txt}[1]    {\ensuremath{\mathrm{#1}}}
\newcommand {\hrefurl}[1]{\href{#1}{\url{#1}}}

\newcommand {\Ref}[1]    {Ref.~\cite{#1}}

\newcommand {\Eq}[1]     {Equation~(\ref{#1})}

\newcommand {\Fig}[1]    {Fig.~\ref{#1}}

\begin{document}

\setcounter{page}{1}

\setlength{\abovecaptionskip}{4pt plus1pt minus1pt}   
\setlength{\belowcaptionskip}{4pt plus1pt minus1pt}   
\setlength{\abovedisplayskip}{6pt plus1pt minus1pt}   
\setlength{\belowdisplayskip}{6pt plus1pt minus1pt}   
\addtolength{\thinmuskip}{-1mu}            
\addtolength{\medmuskip}{-2mu}             
\addtolength{\thickmuskip}{-2mu}           


\title{
The eccentricities of flow - elliptic flow fluctuations and evidence
for transverse localization in the initial state of the matter in
relativistic heavy ion collisions\footnote{Presented at the
International Workshop on Hadron Physics and Properties of High
Baryon Density Matter, Xi'an, China, November 2006.}}

\author{Steven Manly
 for the PHOBOS Collaboration \\
 \mbox{  } \\
{\bf PHOBOS Collaboration } \\
B.Alver$^4$, B.B.Back$^1$, M.D.Baker$^2$, M.Ballintijn$^4$,
D.S.Barton$^2$, R.R.Betts$^6$, R.Bindel$^7$, W.Busza$^4$,
Z.Chai$^2$, V.Chetluru$^6$, E.Garc\'{\i}a$^6$, T.Gburek$^3$,
K.Gulbrandsen$^4$, J.Hamblen$^8$, I.Harnarine$^6$, C.Henderson$^4$,
D.J.Hofman$^6$, R.S.Hollis$^6$, R.Ho\l y\'{n}ski$^3$, B.Holzman$^2$,
A.Iordanova$^6$, J.L.Kane$^4$, P.Kulinich$^4$, C.M.Kuo$^5$,
W.Li$^4$, W.T.Lin$^5$, C.Loizides$^4$, S.Manly$^8$,
A.C.Mignerey$^7$, R.Nouicer$^2$, A.Olszewski$^3$, R.Pak$^2$,
C.Reed$^4$, E.Richardson$^7$, C.Roland$^4$, G.Roland$^4$,
J.Sagerer$^6$, I.Sedykh$^2$, C.E.Smith$^6$, M.A.Stankiewicz$^2$,
P.Steinberg$^2$, G.S.F.Stephans$^4$, A.Sukhanov$^2$, A.Szostak$^2$,
M.B.Tonjes$^7$, A.Trzupek$^3$, G.J.van~Nieuwenhuizen$^4$,
S.S.Vaurynovich$^4$, R.Verdier$^4$, G.I.Veres$^4$, P.Walters$^8$,
E.Wenger$^4$, D.Willhelm$^7$, F.L.H.Wolfs$^8$, B.Wosiek$^3$,
K.Wo\'{z}niak$^3$, S.Wyngaardt$^2$,
B.Wys\l ouch$^4$\\
$^1$ Physics Division, Argonne National Laboratory, Argonne, IL
60439-4843\\ $^2$ Chemistry and C-A Departments, Brookhaven National
Laboratory, Upton, NY 11973-5000\\ $^3$ Institute of Nuclear Physics
PAN, Krak\'{o}w, Poland\\ $^4$ Laboratory for Nuclear Science,
Massachusetts Institute of Technology, Cambridge, MA 02139-4307\\
$^5$ Department of Physics, National Central University, Chung-Li,
Taiwan\\ $^6$ Department of Physics, University of Illinois at
Chicago, Chicago, IL 60607-7059\\ $^7$ Department of Chemistry and
Biochemistry, University of Maryland, College Park, MD 20742\\ $^8$
Department of Physics and Astronomy, University of Rochester,
Rochester, NY 14627\\ }
\date{\today}

\begin{abstract}
Recent measurements of event-by-event elliptic flow in Au+Au
collisions at $\sqrt{s_{\rm NN}}=200$ GeV exhibit large relative
fluctuations of about \mbox{$40$--$50$\%}. The data are well
described by fluctuations in the shape of the initial collision
region, as estimated event-by-event with the participant
eccentricity using Glauber Monte Carlo.  These results, combined
with the demonstrated participant eccentricity scaling of the
elliptic flow across nuclear species, constitute evidence of
transverse granularity in the initial matter production in these
collisions.
\end{abstract}

\maketitle


\section{Introduction}

Elliptic flow~($v_2$) is one of the key observables in the
understanding of the dynamics of heavy ion collisions. The
observation of a significant azimuthal anisotropy in the momentum
and/or spatial distributions of the detected particles relative to
the reaction plane, is direct evidence of interactions between the
initially produced particles in heavy ion collisions. These
interactions must occur at relatively early times, since expansion
of the source rapidly reduces the magnitude of the spatial
asymmetry.

Typically, the connection between the initial and final-state
anisotropy is provided by hydrodynamical models that relate a given,
initial source shape to the distribution of produced particles. In
such calculations, it is common to use smooth, event-averaged,
initial conditions. However, event-by-event fluctuations in the
shape of initial interaction region must not be neglected. As a
means to quantify the effect of initial-state eccentricity
fluctuations, PHOBOS has introduced the ``participant
eccentricity''~\cite{ManlyQM05,PhobosFlowPRL3}.
The
magnitude and shape of $\eps_{part}$  as a function of centrality
were found to be robust to variations of the Glauber parameters.

Fluctuations in the shape of the initial state interaction region
might be expected to be more pronounced for smaller species and for
more peripheral events. This is borne out by the data, where a
striking agreement between the elliptic flow signals in Cu+Cu and
Au+Au data as a function of centrality is obtained when scaled by
the participant eccentricity, $v_2/\langle \eps_{part}
\rangle$~\cite{ManlyQM05,PhobosFlowPRL3}.

Given the obvious importance of fluctuations in the initial state to
the final elliptic flow measurement, PHOBOS recently developed a new
technique to measure dynamical flow fluctuations in our
data~\cite{Alver06,Loizides06}. The new analysis and results
stemming from it are summarized below.

\section{Elliptic flow fluctuations analysis technique and results}

The PHOBOS flow fluctuations analysis extracts the flow signal,
event-by-event, from data in the PHOBOS multiplicity array which
detects a very large fraction of the produced particles over the
pseudorapidity range $|\eta|<5.4$~\cite{WhitePaper}.  The
event-by-event measurement is done using a maximum likelihood fit
with two parameters to the hit information over the full acceptance
of pseudorapidity.  The parameters determined in the maximum
likelihood fit are the observed elliptic flow at midrapidity,
$v_{2}$(0), and the reaction plane angle, $\phi_{0}^{\txt{obs}}$.
The pseudorapidity dependence in the likelihood fit, $v_2(\eta)$, is
parametrized with a triangular shape, {$v^{\rm
tri}_2(\eta)=v_2\,(1-\frac{\abs{\eta}}{6})$}, or alternatively with
a trapezoidal shape, {$v^{\rm trap}_2(\eta) =
\left\{^{v_2\,\txt{if}\,\abs{\eta}<2\,}_{\frac{3}{2}\, v^{\rm
tri}_2(\eta)\,\txt{if}|\eta|>2}\right.$}, where $v_2\equiv v_2(0)$.
Both parametrizations yield good descriptions of the previously
measured~(mean) $v_2(\eta)$ shapes~\cite{PhobosFlowPRC}.  The fit
includes a probability density function that corrects for
non-uniformities in the acceptance of the used sub-detectors.

In order to disentangle known~(mostly statistical) from
unknown~(dynamical) contributions to the measured flow fluctuations,
a detailed knowledge of the detector response is required. A
response function, $K(v_{2}^{\txt{obs}},v_{2}, n)$, is defined as
the distribution of the event-by-event observed elliptic
flow,~$v_{2}^{\txt{obs}}$, for events with constant input flow
value,~$v_2$, and multiplicity,~$n$. This response function is
designed to account for detector deficiencies, as well as for
multiplicity and finite-number fluctuations when mapping the true
distribution of $v_2$ to the distribution of $v_{2}^{\txt{obs}}$. If
$f(v_2)$ is the true $v_2$ distribution for a set of events in a
given centrality class, $f(v_2)$ is related to the distribution of
$v_2^{\txt{obs}}$, $g(v_{2}^{\txt{obs}})$, by
\begin{equation}
 g(v_{2}^{obs}) = \int K(v_{2}^{\txt{obs}},v_2,n) \, f(v_2)
 \, N(n) \, \txt{d}v_2  \, \txt{d}n\,,
 \label{eqkernel}
\end{equation}
where $N(n)$ is the multiplicity distribution of the events in the
given set of events.

To obtain the kernel in bins of $v_2$ and $n$ with enough precision,
would require on the order of 100 million MC events. Instead the
kernel can be parametrized, allowing the use of about 1.5\% of that
statistics to reach the required precision. For a perfect detector,
the response is given by Eq.~(A13) from \Ref{Ollitrault1992}~(with
$\alpha \rightarrow v_2^{\txt{obs}}$, $\overline{\alpha} \rightarrow
v_2$ and $M \rightarrow n$). In practice, however, it turns out that
$v_2$ is suppressed, with the suppression dependent on $n$, and that
the resolution~($\sigma$) has a constant background contribution.
With $v_{2}^{\txt{sup}}=(A\,n+B)v_{2}$ and $\sigma = C/\sqrt{n} +
D$, this leads to
\begin{multline}
 K(v_{2}^{\txt{obs}},v_{2},n) = \frac{v_2^{\txt{obs}}}{\sigma^2} \,
     \exp  \left(-\frac{\left(v_2^{\txt{obs}}\right)^2+
     \left(v_2^{\txt{sup}}\right)^2}{2\sigma^2}\right)\\
     I_{0} \left(-\frac{v_{2}^{\txt{obs}}v_2^{\txt{sup}}}{\sigma^2}\right)
\end{multline}
where $I_0$ is a modified Bessel function. The four unknown
parameters~($A,B,C,D$) are obtained using the modified HIJING
samples.

In order to determine the mean and variance of the true $v_2$
distribution, $f(v_2)$, and extract the fluctuations, we assume a
Gaussian distribution for $f(v_2)$, with two parameters, $\langle
v_{2}\rangle$ and $\sigma_{v_2}$. For given values of $\langle
v_{2}\rangle$ and $\sigma_{v_2}$, it is possible to take the
integral in \Eq{eqkernel} to obtain the expected distribution,
\mbox{$g_{\txt{exp}}(v_2^{\txt{obs}}|\langle
v_{2}\rangle,\sigma_{v_2})$}. By comparing the expected and observed
distributions, the values for $\langle v_2 \rangle$ and
$\sigma_{v_2}$ are found by a maximum likelihood fit.

\begin{figure}[h]
\centerline{ \epsfig{file=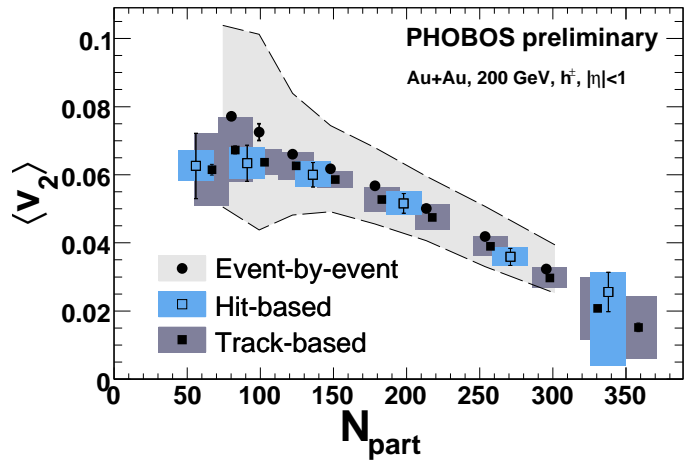,width=9.0cm} }
\caption{ Mean flow, $\av{v_2}$,
          as a function of centrality,
          for $\snn=200$~GeV collisions at mid-rapidity, measured
          by the event-by-event analysis, compared to the published
          results obtained with the event-averaged, sub-event based
          technique using hits or tracks~\cite{PhobosFlowPRC}.
 } \label{fig1}
\end{figure}

The analysis chain is applied to the $\snn=200$~GeV Au+Au data set
from Run~4 in bins of centrality. The results~($\langle
v_{2}\rangle$ and $\sigma_{v_2}$) are obtained separately for
triangular and trapezoidal $v_2(\eta)$ shape and averaged over 10
bins of collision vertex~($2$~cm width). The systematic errors~(all
sources added in quadrature) are estimated by including variations
from different vertex and $\phi_0^{\txt{obs}}$ bins and changes
introduced by the triangular and trapezoidal $v_2(\eta)$ shapes.
Since the functional form of the true distribution is unknown, also
differences arising from a flat rather than a Gaussian ansatz for
$f(v_2)$ are included. Furthermore, we have performed extensive
studies of the analysis response to MC samples, prepared to match
the observed multiplicity distribution and $\langle v_2\rangle$, in
bins of centrality and known input values of fluctuations~(including
zero). The discrepancies of the obtained to the input fluctuations
in the vicinity~($\pm20$\%) of the measured values per centrality
bin are an additional source of the systematic error. Their
contribution becomes increasingly important for lower values of
$v_2$, which is the dominant reason to exclude the \mbox{$0$--$6$\%}
most central bin from the analysis. \Fig{fig1} shows the mean flow
results obtained with the event-by-event analysis projected to
mid-rapidity by $\av{v_2}=0.5\left(\frac{11}{12} \av{v_2^{\rm tri}}
+ \av{v_2^{\rm trap}}\right)$, which agree well, within scale
errors, with the published results of the event-averaged sub-event
based technique using hits or tracks~\cite{PhobosFlowPRC}. This
constitutes an important, completely independent verification of the
event-by-event analysis. \Fig{fig2} presents the relative flow
fluctuations,~$\sigma_{v_2}/\av{v_2}$. In the ratio most of the
forementioned systematic errors scale out. Large relative
fluctuations of about \mbox{$40$--$50$\%} are observed, with
relative little centrality dependence.  The PHOBOS results presented
here are consistent with a recent, preliminary measurement of
elliptic flow fluctuations by the STAR
collaboration~\cite{sorensen}.

\section{Fluctuations expected from the participant eccentricity model}

\begin{figure}[h]
\centerline{
\epsfig{file=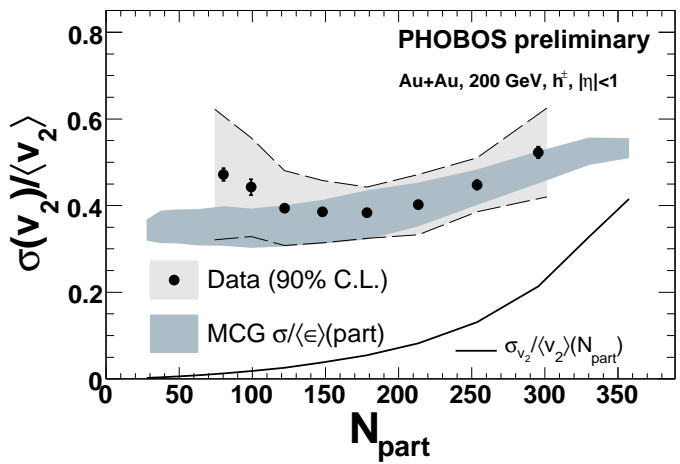,width=9.0cm}
} \caption{ Relative flow fluctuations,~$\sigma_{v_2}/\av{v_2}$,
          as a function of centrality,
          for \mbox{$\snn=200$} GeV collisions at mid-rapidity,
          compared to the prediction, $\sigma_{\eps}/\av{\eps}$,
          from the participant eccentricity, and to an estimate of
          $N_{\rm part}$ induced fluctuations
          using a fit of $\av{v_2}(N_{\rm part})$
 } \label{fig2}
\end{figure}

The participant eccentricity picture accounts for nucleon-position
fluctuations in the participating nucleon distributions by
calculating the eccentricity, event-by-event, with respect to the
principal axes of the overlap ellipse in a MC Glauber~(MCG)
simulation. In a hydrodynamical scenario, such fluctuations in the
shape of the initial collision region would lead naturally to
corresponding fluctuations in the elliptic flow signal. To estimate
their magnitude, it is assumed that $v_2\propto\eps$ event-by-event.
This leads to ${\sigma_{v_2}}/{\langle v_2 \rangle} =
{\sigma_{\eps}}/{\langle \eps \rangle}$, where
$\sigma_{v_2}$~($\sigma_{\eps}$) is the standard deviation of the
event-by-event distribution $v_2$~($\eps$), provided there are no
other sources of elliptic flow fluctuations. Neglecting all other
sources of elliptic flow fluctuations, the participant eccentricity
MCG simulation predicts relative fluctuations~($\sigma_{v_2}/\langle
v_2 \rangle$) of \mbox{$35$--$50$\%} in Au+Au collisions at
$\snn=200$~GeV.  The prediction is shown in \Fig{fig2} as a function
of $N_{\rm part}$.
The data are well described by the prediction,
$\sigma_{\eps}/\av{\eps}$, from the participant eccentricity
obtained in MCG simulations~(to determine the 90\% confidence level
band shown in \Fig{fig2}, the Glauber parameters were varied within
reasonable limits as described in~\Ref{PhobosFlowPRL3}). The
contribution from $N_{\rm part}$ fluctuations, estimated using a fit
to the $\av{v_2}$ data and the known $N_{\rm part}$ distributions
from the PHOBOS centrality trigger studies, is neglible in the
measured centrality range.

\section{Conclusions}
Recent results on event-by-event elliptic flow fluctuations in Au+Au
collisions at \mbox{$\sqrt{s_{\rm NN}}=200$} GeV exhibit large
relative fluctuations of about \mbox{$40$--$50$\%}, relatively
independent of centrality. The new data are well described by
fluctuations in the shape of the initial collision region, as
predicted with the participant eccentricity using MCG simulations.

These results substantiate conclusions from previous studies by
PHOBOS on the relevance of such event-by-event fluctuations for the
elliptic flow across nuclear species.  The initial-state geometry
seems to drive the hydrodynamic evolution of the system, not only on
average, but event-by-event. The success of the participant
eccentricity model in describing both the geometric scaling and
fluctuations of the elliptic flow is consistent with the matter
present in the initial stage of relativistic heavy ion collisions
being created with a transverse granularity similar to that of the
participating nucleons.

\section*{Acknowledgments}
%
%
%
%
This work was partially supported by U.S. DOE grants
DE-AC02-98CH10886, DE-FG02-93ER40802,
DE-FG02-94ER40818,  
DE-FG02-94ER40865, DE-FG02-99ER41099, and DE-AC02-06CH11357, by U.S.
NSF grants 9603486, 
0072204,            
and 0245011,        
by Polish KBN grant 1-P03B-062-27(2004-2007), by NSC of Taiwan
Contract NSC 89-2112-M-008-024, and by Hungarian OTKA grant (F
049823).

\end{document}